\documentclass[aps, prl, preprint, showpacs, bibliography]{revtex4-1}
\usepackage{amssymb, amsmath, bm, graphicx, color}
\usepackage{hyperref}
\begin{document}

%\preprint{}

\title{Performance analysis of Schottky-junction-based near-field thermophotovoltaic system}


\author{Jaeman Song, Mikyung Lim, Seung S. Lee, and Bong Jae Lee}
\email{bongjae.lee@kaist.ac.kr}
%\thanks{}
\affiliation{Department of Mechanical Engineering,
Korea Advanced Institute of Science and Technology, Daejeon 34141, South Korea}
%option in \documentclass).\noaffiliation is required (may also be

\date{\today}

\begin{abstract}
Numerous studies have reported performance enhancement of a thermophotovoltaic (TPV) system when an emitter is separated by nanoscale gaps from a TPV cell. Although a $p$-$n$-junction-based TPV cell has been widely used for the near-field TPV system, a Schottky-junction-based near-field TPV system has drawn attention recently with the advantage of the easy fabrication. However, existing studies mostly focused on the generated photocurrent only in the metal side due to the fact that required energy for the metal-side photocurrent (i.e., Schottky barrier height) is smaller than the bandgap energy. Here, we suggest the precise performance analysis model for the Schottky-junction-based near-field TPV system, including photocurrent generation on the semiconductor side by considering the transport of minority carriers within the semiconductor. It is found that most of the total photocurrent in the Schottky-junction-based near-field TPV system is generated in the semiconductor side. We also demonstrate that further enhancement in the photocurrent generation can be achieved by re-absorbing the usable photon energy in the metal with the help of a backside reflector. The present work will provide a design guideline for the Schottky-junction-based near-field TPV system taking into account three types of photocurrents.
\end{abstract}
%\keywords{}
\maketitle

\section{Introduction}
Thermophotovoltaic (TPV) system has widely received attention as a promising energy conversion device that directly converts the absorbed thermal radiation into the electricity with advantages of the quiet operation and potential of the miniaturization \cite{basu2007microscale, park2013fundamentals, tervo2018near}. As one way to further improve the performance of the TPV system, the near-field radiation has been applied (i.e., near-field TPV system). When the vacuum gap distance between the emitter and the receiver is smaller than the thermal characteristic wavelength determined by the Wien's displacement law \cite{bergman2011fundamentals}, the magnitude of radiative heat transfer can surpass the blackbody limit through the coupling of evanescent waves (i.e., photon tunneling) \cite{basu2016near}. In fact, it has already been shown that performance enhancement of the TPV system can be achieved by reducing the gap between the emitter and the TPV cell \cite{park2008performance, francoeur2011thermal, svetovoy2014graphene, bernardi2015impacts, lim2015graphene, tong2015thin, jin2016hyperbolic, st2017hot, lim2018optimization}.

One representative structure of the TPV cell is a \textit{p-n} junction semiconductor, and there have been extensive theoretical works regarding the \textit{p-n}-junction-based near-field TPV system \cite{park2008performance, francoeur2011thermal, bernardi2015impacts, lim2015graphene, tong2015thin, jin2016hyperbolic, karalis2016squeezing, lim2018optimization}. Under illumination of photons with wavelength shorter than the wavelength corresponding to the bandgap of the semiconductor, electron-hole pairs are generated inside the semiconductor, diffused toward the depletion region and separated by built-in electric field across the region, which yields the net current flow (i.e., photocurrent generation) \cite{sze2006physics}. Recently, the enhanced performance of the near-field TPV system was experimentally demonstrated using the \textit{p-n}-junction-based TPV cell made of InAs \cite{fiorino2018nanogap}; however, challenges in fabricating the perfectly flat \textit{p-n} junction TPV cell as well as its high cost and high bandgap energy make experimental demonstration rather limited \cite{dimatteo2001enhanced, st2017hot, fiorino2018nanogap}.

Alternatively, the Schottky junction can also be utilized as a TPV system \cite{svetovoy2014graphene, st2017hot, yang2018observing,vongsoasup2018effects}, and it can be formed when a metal, which has a greater work function than the electron affinity of a semiconductor, makes contact with the semiconductor \cite{sze2006physics}. Comparing to the \textit{p-n} junction, the Schottky junction has an advantage in the simple fabrication process because there is no need for high temperature and expensive fabrication steps like diffusion and annealing \cite{godfrey1979655, hezel1997recent}. Furthermore, through the simplified fabrication process, surface conditions (e.g., cleanliness, roughness, and planarity) of the TPV cell can be well-controlled such that the nanoscale vacuum gap between the emitter and the TPV cell can be maintained relatively easier than the case of \textit{p-n}-junction-based TPV cell. Further, the Schottky-junction-based TPV cell is also able to produce the metal-side photocurrent (i.e., internal emission photocurrent) by absorbing photon with less energy than bandgap energy of semiconductor, making it advantageous when the bandgap energy is greater than that of the absorbed radiation.

Accordingly, all the previous studies on the Schottky-junction-based near-field TPV system have mainly analyzed the electrical power generation from the metal side because they employed the semiconductor with high bandgap energy (e.g., silicon) \cite{svetovoy2014graphene, st2017hot, yang2018observing}. However, significance of the photocurrent generation from the semiconductor side of a Schottky-junction-based TPV cell has been emphasized continuously \cite{liou1996simple, sze2006physics, farhat2017plasmonically, johnston2008efficient}, and even several studies have suggested that in order to obtain the high photocurrent density, the absorption of the incident photon by the metal should be minimized because the quantum efficiency of the internal emission photocurrent is marginal \cite{sze2006physics}. Although recent works have demonstrated that the quantum efficiency of the internal emission photocurrent can be improved depending on the various factors, such as the fabrication condition, the geometry of the metallic layer, and surface plasmon polariton \cite{sze2006physics, scales2010thin, sundararaman2014theoretical, bernardi2015theory, blandre2018limit}, it is crucial to contemplate the photocurrent generation from the semiconductor side when designing a near-field TPV system. For example, if we employ a low-bandgap semiconductor (e.g., GaAs, InSb, or InAs) and an appropriate metal for the Schottky junction, the generated photocurrent from the semiconductor side can also contribute to the total generated photocurrent, meaning that there is a room for significantly improving the performance of the Schottky-junction-based near-field TPV system.

In this work, we propose a model for predicting the performance of Schottky-junction-based near-field TPV system and analyze the performance accordingly with the basic configuration. Tungsten is used as an emitter and nickel-$n$-doped GaSb Schottky-junction-based TPV cell is selected as a receiver. Photocurrents generated from either sides will be compared and a proper thickness of the metal for an enhanced performance will be discussed. Furthermore, an additional performance improvement will be achieved by introducing the backside reflector, which can increase the photocurrent generation from the metal side.

%%%%%%%%%%
\begin{figure}[!t]
\centering\includegraphics[width=0.45\textwidth]{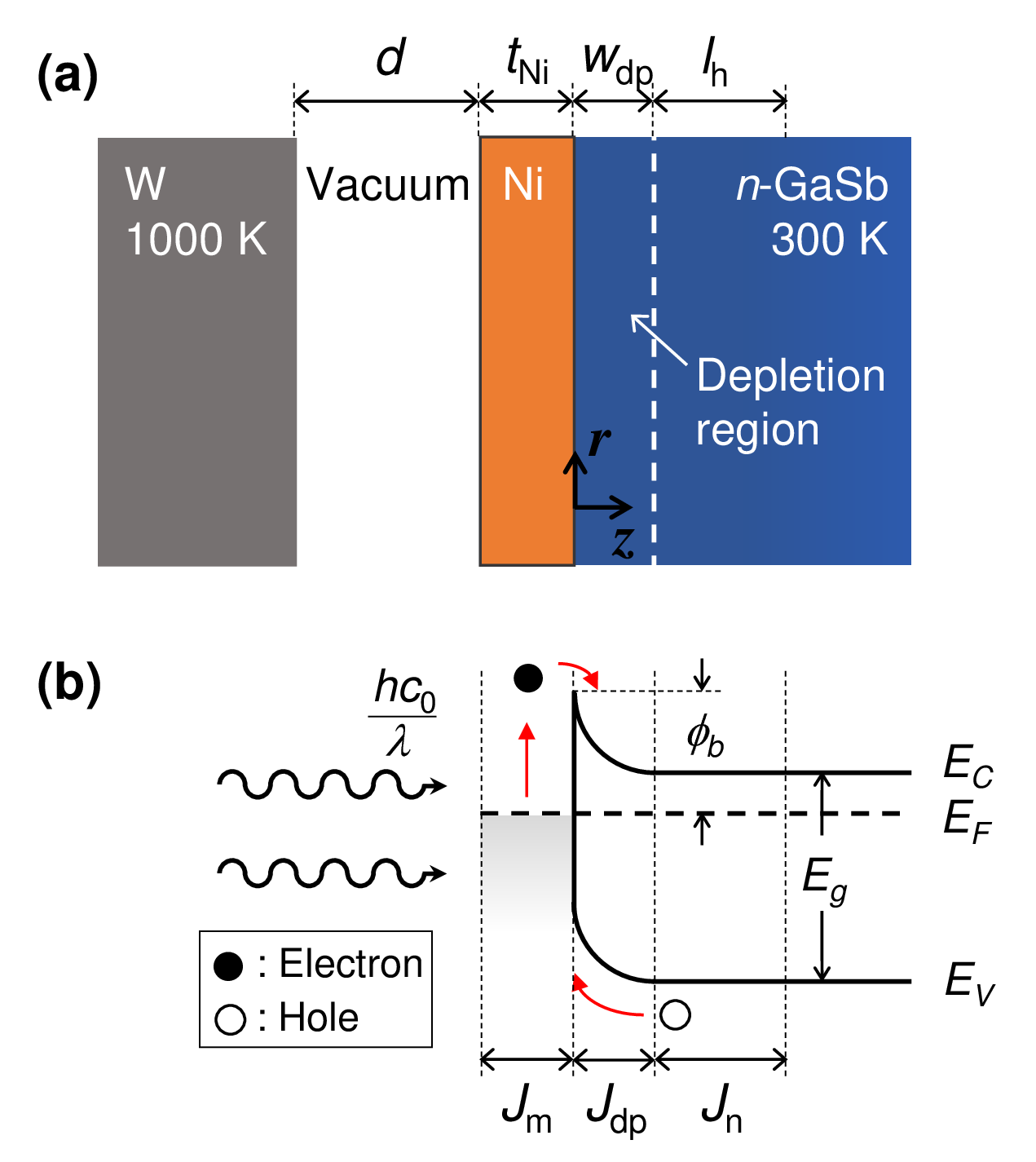}
\caption{(a) Schematic of Schottky-junction-based near-field TPV system consisting of a tungsten emitter and a nickel deposited \textit{n}-doped GaSb TPV cell. (b) Energy-band diagram for the Schottky-junction-based TPV cell.}
\label{Fig:1}
\end{figure}
%%%%%%%

%%==========================================
\section{Model and method}

Figure \ref{Fig:1}a illustrates a schematic of the Schottky-junction-based near-field TPV system. As an emitter, tungsten is maintained at 1000 K. The Schottky-junction-based TPV cell maintained at 300 K is located with a vacuum gap distance, $d$ from the emitter. In order to make the Schottky junction, 5-nm-thick nickel layer is deposited on the $n$-doped GaSb, which is the verified configuration in the recent experimental work  \cite{vongsoasup2018effects}. 

In order to calculate the absorbed net radiative heat flux in each layer of the Schottky-junction-based TPV cell, formulations for the near-field thermal radiation between multilayered structures considering both forward and backward waves in each layer are employed \cite{park2008performance, francoeur2009solution}. The net radiative heat flux can be expressed as $q''_{net}=\int_0^{\infty} d\lambda\ q''_{\lambda, net} = \int_0^{\infty} d\lambda\ \int_0^{\infty} S_{\beta, \lambda}(\beta, \lambda) d\beta$, where $\lambda$ refers to vacuum wavelength, $\beta$ is the parallel wavevector component and the expression for the $S_{\beta, \lambda}(\beta, \lambda)$ can be found in Refs.\ \cite{park2008performance, francoeur2009solution}. The dielectric function of tungsten for $\lambda<10$ $\mu\text{m}$ is obtained from the tabular data in Ref.\ \cite{palik1998handbook} and that for $\lambda\geq 10$ $\mu\text{m}$ is estimated from the Drude model \cite{ordal1985optical}. For the nickel and gold, the dielectric functions are obtained from the Lorentz Drude model \cite{rakic1998optical}. The bandgap energy, $E_g$ of the \textit{n}-doped GaSb is calculated as 0.72 eV using Varshini\textquoteright s equation \cite{gonzalez2006modeling}, and its dielectric function is estimated using the model suggested by Adachi \cite{adachi1989optical} (for $E\geq E_g$) and Patrini \cite{patrini1997optical} (for $E<E_g$).

%%==========================================
\subsection{Power generation in the Schottky-junction-based TPV cell}
In order to make the Schottky junction, the work function of the metal and the electron affinity of the semiconductor should be considered first. The work function of a metal, $\phi_m$ is defined as the energy difference between the vacuum level and the Fermi level. The electron affinity of a semiconductor, $\chi$ is the energy difference between the vacuum level and the conduction band. When a high-work-function metal and a low-electron-affinity semiconductor make a contact, the ideal Schottky barrier height, $\phi_b=\phi_m-\chi$ is formed after lining up the the Fermi levels on both materials \cite{sze2006physics}. The Schottky-junction-based TPV cell used in this work (i.e., Ni / $n$-doped GaSb) has $\phi_b$ = 0.4976 eV \cite{vongsoasup2018effects}. On the other hand, if a low-work-function metal and a high-electron-affinity semiconductor are used to make a contact, then, an ohmic junction is formed, which can be used for an electrode or a backside reflector because of its low electrical junction resistance \cite{sze2006physics}.

When a metal and a semiconductor make a contact, a depletion region is formed adjacent to the metal layer so that the built-in electric field is formed (refer to Fig.\ \ref{Fig:1}a). The width of depletion region, $w_{dp}$ is determined using Poisson’s equation under the approximation of abrupt junction \cite{rhoderick1982metal, pierret1996semiconductor, sze2006physics}:
\begin{equation}\label{Eq:1}
w_{dp}=\sqrt{\frac{2\varepsilon_s}{eN_D}\left(\psi_{bi}-V_f\right)}
\end{equation}
where $\varepsilon_s=1.390\times10^{-10}$ Fm$^{-1}$ is the static permittivity of the GaSb \cite{gonzalez2006modeling}, $N_D=1\times10^{17}$ cm$^{-3}$ is the $n$-doping concentration, $e$ is the electron charge, and $V_f$ is the forward bias. The built-in potential, $\psi_{bi}$ is obtained from \cite{pierret1996semiconductor, sze2006physics}
\begin{equation}\label{Eq:2}
\psi_{bi}=\phi_b-\left(E_C-E_F\right)
\end{equation}
In Eq.\ (\ref{Eq:2}), for the non-degenerate semiconductor, the difference between conduction level, $E_C$ and Fermi level, $E_F$ is calculated by $E_C-E_F=k_BT\text{ln}\left(\frac{N_C}{N_D}\right)$, where $k_B$ is the Boltzmann constant and $T$ is the temperature of the TPV cell. Because the effective density of states in the conduction band, $N_C$ = $2.806\times10^{17}$ cm$^{-3}$ is larger than $N_D$, the considered $n$-doped GaSb can be regarded as the non-degenerate semiconductor \cite{sze2006physics}. We also assume that $n$-doped GaSb is fully ionized given that the thermal energy $k_BT$ is comparable to the ionization energy at 300 K \cite{sze2006physics}. With the assumption that variation of the width of the depletion region by forward bias can be neglected (i.e., $V_f=0$), Eq.\ (\ref{Eq:1}) yields 90.4-nm-thick depletion region for the considered nickel-$n$-doped GaSb junction. We further assume that the depletion region is formed in the $n$-GaSb side and treat the rest part of the $n$-GaSb as neutral region.

As illustrated in the energy-band diagram of the Schottky-junction-based TPV system (refer to Fig.\ \ref{Fig:1}b), electron transport from the metal to the semiconductor can contribute to the photocurrent generation. In addition, movement of holes from the semiconductor to the metal can also generate the photocurrent. The photocurrents generated in each layer are noted as $J_m$ (i.e., in the metal layer), $J_{dp}$ (i.e., in the depletion region), and $J_n$ (i.e., in the neutral region), respectively.

When the photon whose energy is greater than the Schottky barrier height ($\phi_b$) is absorbed in the metal layer, the photo-excited electron surmounts the Schottky barrier height and are collected in the semiconductor. These photo-excited electrons produce the spectral internal emission photocurrent, $J_{m,\lambda}$ that can be computed using the quantum efficiency, $\eta_{\lambda,m}$ considering multiple reflections of the hot electron in the thin metal film \cite{scales2010thin}. If the thickness of metal is thinner than the electron mean free path (e.g., 5.87 nm for Ni \cite{gall2016electron}), it is regarded as the thin film. The spectral internal emission photocurrent, $J_{m,\lambda}$ is then calculated using the following equation:
\begin{equation}\label{Eq:3}
J_{m,\lambda}\left(\lambda\right)=\eta_{\lambda,m}e\frac{q ''_{\lambda,m}}{hc_0/\lambda} \quad \text{with} \quad  \frac{hc_0}{\lambda}\geq \phi_b
\end{equation}
where $q ''_{\lambda,m}$ is the spectral radiative heat flux absorbed in the metal, $h$ is the Planck constant, and $c_0$ is the speed of light in vacuum.

When the photon whose energy is greater than the bandgap energy ($E_g$) passes through the metal layer and is absorbed in the semiconductor, electron-hole pair is generated inside the semiconductor. In such situation, the diffused and swept minority carriers produce diffusion and drift photocurrents, respectively \cite{park2008performance, francoeur2011thermal, lim2015graphene, sze2006physics}. In the neutral region of GaSb, within the diffusion length of holes ($l_h$), the diffusion photocurrent is generated when the minority carrier diffuses and reaches to the depletion region. The following one-dimensional steady state continuity equation is used to calculate the concentration of diffused minority carrier $n_h(z,\lambda)$ considering the recombination \cite{park2008performance, francoeur2011thermal, lim2015graphene}:
\begin{equation}\label{Eq:4}
D_h\frac{d^2}{dz^2} \Bigg [ n_h(z,\lambda)-n^0_h \Bigg ] -\frac{n_h(z,\lambda)-n^0_h}{\tau_h}+\dot{g}(z,\lambda)=0
\end{equation}
where $\dot{g}(z,\lambda)=-\frac{dQ_\lambda(z,\lambda)}{dz} \times \frac{\lambda}{hc_0}$ is the spectral photogeneration rate of electron-hole pairs with $Q_\lambda(z,\lambda)= \int_0^{\infty} S_{\beta, \lambda,sc}(\beta, \lambda)\  e^{-2\Im(k_{z,sc})z}\ d\beta$ being the net radiative heat flux inside the TPV cell. Note that $S_{\beta, \lambda,sc}$ indicates the $S_{\beta, \lambda}$ absorbed in the semi-infinite semiconductor and $k_{z,sc}$ is the $z$-component wavevector while the subscript `sc' represents the semiconductor.  In Eq.\ (\ref{Eq:4}), $n^0_h$ is the equilibrium hole concentration, $D_h$ is the diffusion coefficient, and $\tau_h$ is the life time of holes. Here, $n_h(z,\lambda)-n^0_h$ is solved by employing the semi-analytic method \cite{lim2015graphene} with the following boundary conditions: (i) the hole concentration is assumed to be equilibrium at the edge of the depletion region (i.e., $n_h=n^0_h$ at $z=w_{dp}$); and (ii) surface recombination is neglected in the region farther than the diffusion length from the depletion region edge (i.e., $\left.D_h\frac{dn_h}{dz}\right|_{z=w_{dp}+3l_h}=0$). The spectral hole diffusion photocurrent can then be calculated by
\begin{equation} \label{Eq:5}
\left.J_{n,\lambda}\left(\lambda\right)=-eD_h\frac{dn_h(z,\lambda)}{dz}\right|_{z=w_{dp}} \quad \text{with} \quad \frac{hc_0}{\lambda}\geq E_g    
\end{equation}
The electrical properties (e.g., diffusion coefficient and lifetime of hole) of $n$-doped GaSb are taken from Ref.\ \cite{gonzalez2006modeling} as $D_h$ = 19 cm$^2$s$^{-1}$ and  $\tau_h = 24.4$ ns when $N_D = 1\times10^{17}$ cm$^{-3}$ . With those values, the hole diffusion length is estimated to be 7 $\mu$m from $l_h=\sqrt{D_h\tau_h}$ \cite{park2008performance}.

In the depletion region, because the minority carriers (i.e., holes for the $n$ type) are swept by the built-in electric field, the spectral drift photocurrent is assumed to be generated without any recombination (i.e., quantum efficiency of 100\%). It can be expressed by
\begin{equation}\label{Eq:7}
J_{dp,\lambda}\left(\lambda\right)=e\frac{Q_\lambda(0,\lambda)-Q_\lambda(w_{dp},\lambda)}{hc_0/\lambda} \quad \text{with} \quad  \frac{hc_0}{\lambda}\geq E_g
\end{equation}
where $Q_\lambda(0,\lambda)-Q_\lambda(w_{dp},\lambda)$ represents the spectral radiative heat flux absorbed in the depletion region.

Finally, the spectral photocurrent density, $J_{ph,\lambda}\left(\lambda\right)$ is calculated by the summation of the generated spectral photocurrents in each layer, i.e., $J_{ph,\lambda}\left(\lambda\right)=J_{m,\lambda}\left(\lambda\right)+J_{dp,\lambda}\left(\lambda\right)+J_{n,\lambda}\left(\lambda\right)$, and the total photocurrent $J_{ph}$ can be obtained from the integration of the spectral photocurrent: $J_{ph}=\int J_{ph,\lambda}(\lambda) d\lambda$. Under illumination, the current-voltage characteristics of the Schottky TPV cell can be defined by subtracting the dark current from the total photocurrent $J_{ph}$: 
\begin{equation}\label{Eq:8}
J=J_{ph}-A^*T^2 \exp \left(-\frac{\phi_b}{k_BT}\right) \left [ \exp \left(\frac{eV}{k_BT}\right)-1\right ]
\end{equation}
where $A^*=5.16\times10^4$ Am$^{-2}$K$^{-2}$ is the Richardson constant of the $n$-doped GaSb \cite{vongsoasup2018effects}. The dark current of the Schottky TPV cell expressed as the second term of the right side of Eq.\ (\ref{Eq:8}) is obtained from the current-voltage characteristics of the Schottky diode considering the thermionic emission of carriers \cite{sze2006physics}. The maximum electrical power output of the TPV system is calculated as the maximum product of current and voltage related with Eq.\ (\ref{Eq:8}).

%%==========================================
\section{Results and discussion}

%%%%%%%%%%
\begin{figure}[!b]
\centering\includegraphics[width=0.95\textwidth]{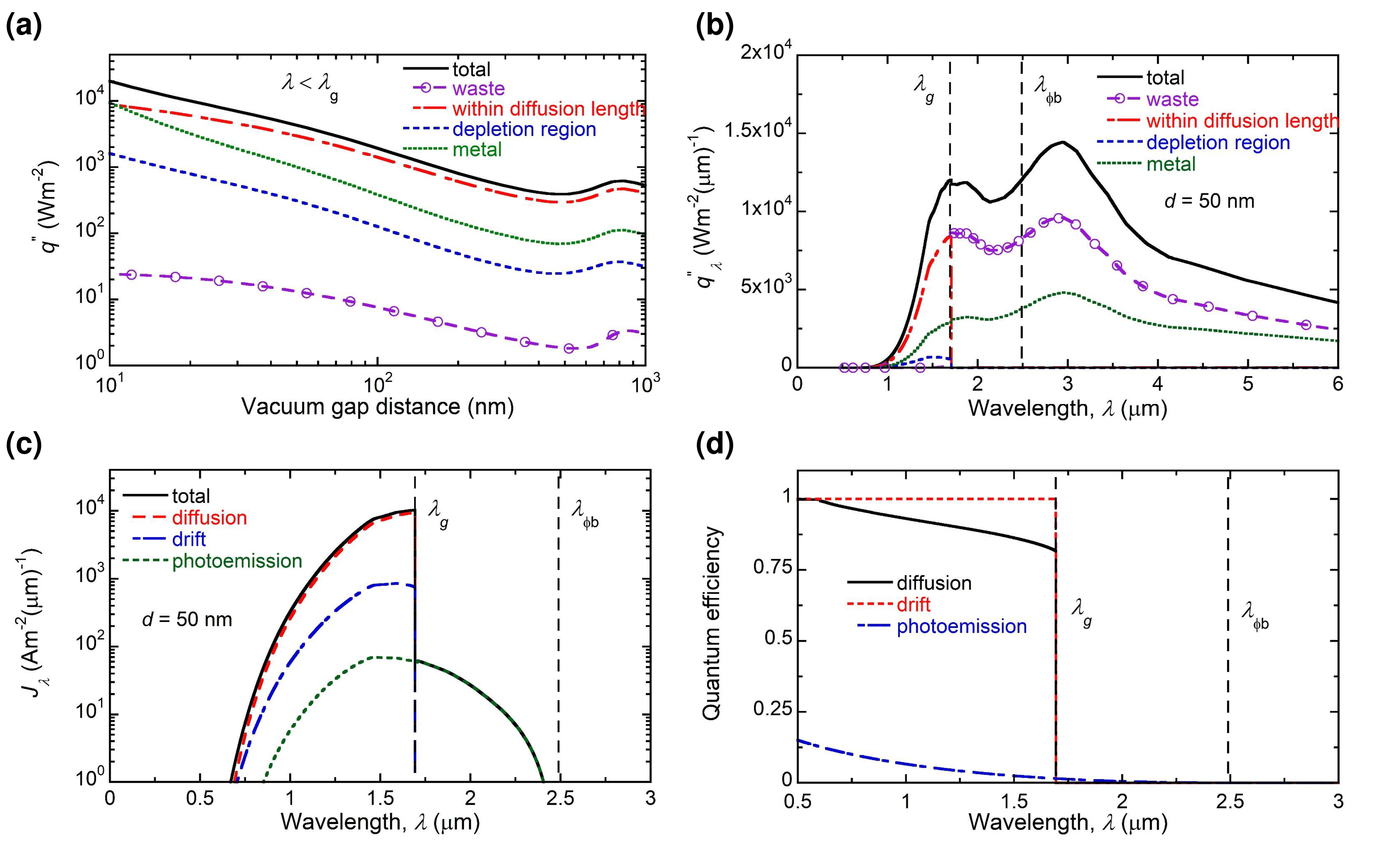}
\caption{(a) Absorbed radiative heat flux with $\lambda<\lambda_g$ in each layer of the Schottky TPV cell. (b) Spectral heat flux, $q^{\prime\prime}_{\lambda}$ when $d=50$ nm. (c) Spectral photocurrent, $J_{\lambda}$ when $d=50$ nm. (d) Quantum efficiency of three generated photocurrents in the Schottky TPV cell.}
\label{Fig:2}
\end{figure}
%%%%%%%

Let us first consider the amount of radiative heat flux absorbed in each layer of the Schottky TPV cell. Figure \ref{Fig:2}a shows the absorbed radiative heat flux with $\lambda < \lambda_g$ in each layer with respect to the vacuum gap distance, $d$. Outside the hole diffusion length of the neutral region of TPV cell, the generated electron-hole pair rarely contributes to the diffusion photocurrent; thus, it is simply noted as `waste' in  Fig.\ \ref{Fig:2}. It can be seen from Fig.\ \ref{Fig:2}a that the 5-nm-thick nickel layer absorbed more radiative energy than the 90.4-nm-thick depletion region for all considered $d$ values. However, the absorbed radiative heat flux in the neutral region within the diffusion length of GaSb is greater than that in the nickel film. Magnitudes of the absorbed radiative heat flux in the metal film and in the neutral region within the diffusion length become comparable only at the extremely small vacuum gap (i.e., $d =10$ nm) as a result of the short radiation penetration depth at such small gap \cite{basu2009ultrasmall, song2015enhancement}. Thus, it can be inferred that the photocurrent generation in the semiconductor side cannot be neglected unless the thickness of metal film is much thicker than the radiation penetration depth at moderate vacuum gap distances of $d>50$ nm.

Figure \ref{Fig:2}b shows the absorbed spectral heat flux in each layer at $d=50$ nm. Here, we consider the vacuum gap distance of $d$ = 50 nm according to the recently reported near-field TPV experiment \cite{fiorino2018nanogap}. For $\lambda> \lambda_g$, the radiative heat flux is rarely absorbed in the depletion region nor in the neutral region within the diffusion length, suggesting that most of the absorbed radiation in the semi-infinite GaSb semiconductor for this spectral regime is simply wasted. On the other hand, nickel layer absorbs a wide spectral range of the radiative heat flux, and the corresponding internal emission photocurrent ($J_{m,\lambda}$) can be produced even at wavelengths longer than the bandgap wavelength as long as the absorbed photon has greater energy than the Schottky barrier height  (i.e., $\lambda<\lambda_{\phi_b}$). Although the absorbed radiation with energy greater than bandgap energy is smaller for the metal side (i.e., approximately 20\% of the total incident radiative heat flux), amount of the useful energy for the metal layer (i.e., $\lambda<\lambda_{\phi_b}$) and that for the neutral region within the diffusion length (i.e., $\lambda < \lambda_g$) are comparable with each other.

Figure \ref{Fig:2}c shows the log-scaled spectral photocurrent density generated in the three different layers: neutral region (i.e., diffusion photocurrent), depletion region (i.e., drift photocurrent), and metal layer (i.e., internal emission photocurrent) at $d$ = 50 nm. It can be clearly seen that the internal emission photocurrent can be generated in the wider spectral range than the drift and diffusion photocurrents. However, $J_{m,\lambda}$ accounts for only about 1.4\% of $J_{ph,\lambda}$ although the absorbed useful radiation in the metal side is comparable to that in the neutral region within the diffusion length. This is due to the low quantum efficiency of the internal emission photocurrent compared to that of drift or diffusion photocurrent in the semiconductor, as can be seen in Fig.\ \ref{Fig:2}d. As mentioned in the previous section, the drift quantum efficiency is assumed to be 100\%, and the diffusion quantum efficiency, which is obtained from $\eta_{n,\lambda}=J_{n,\lambda} \times \frac{hc_0}{e\lambda} \times \frac{1}{Q_\lambda(w_{dp},\lambda)-Q_\lambda(w_{dp}+l_h,\lambda)}$, is larger than 0.8 for  $\lambda < \lambda_g$. However, $\eta_{m,\lambda}<0.2$ in the wavelength of interest ($\lambda < \lambda_{\phi_b}$) and becomes almost negligible in $\lambda_g < \lambda <\lambda_{\phi_b}$, which makes the internal emission photocurrent marginal compared to the drift and diffusion photocurrents (see Fig.\ \ref{Fig:2}c). Consequently, the photocurrent generated in the semiconductor cannot be neglected as previously done in Ref.\ \cite{vongsoasup2018effects}. For enhancing the power output of the considered Schottky-junction-based near-field TPV system, using thinner nickel layer would be beneficial. 

%%%%%%%%%%
\begin{figure}[!b]
\centering\includegraphics[width=0.5\textwidth]{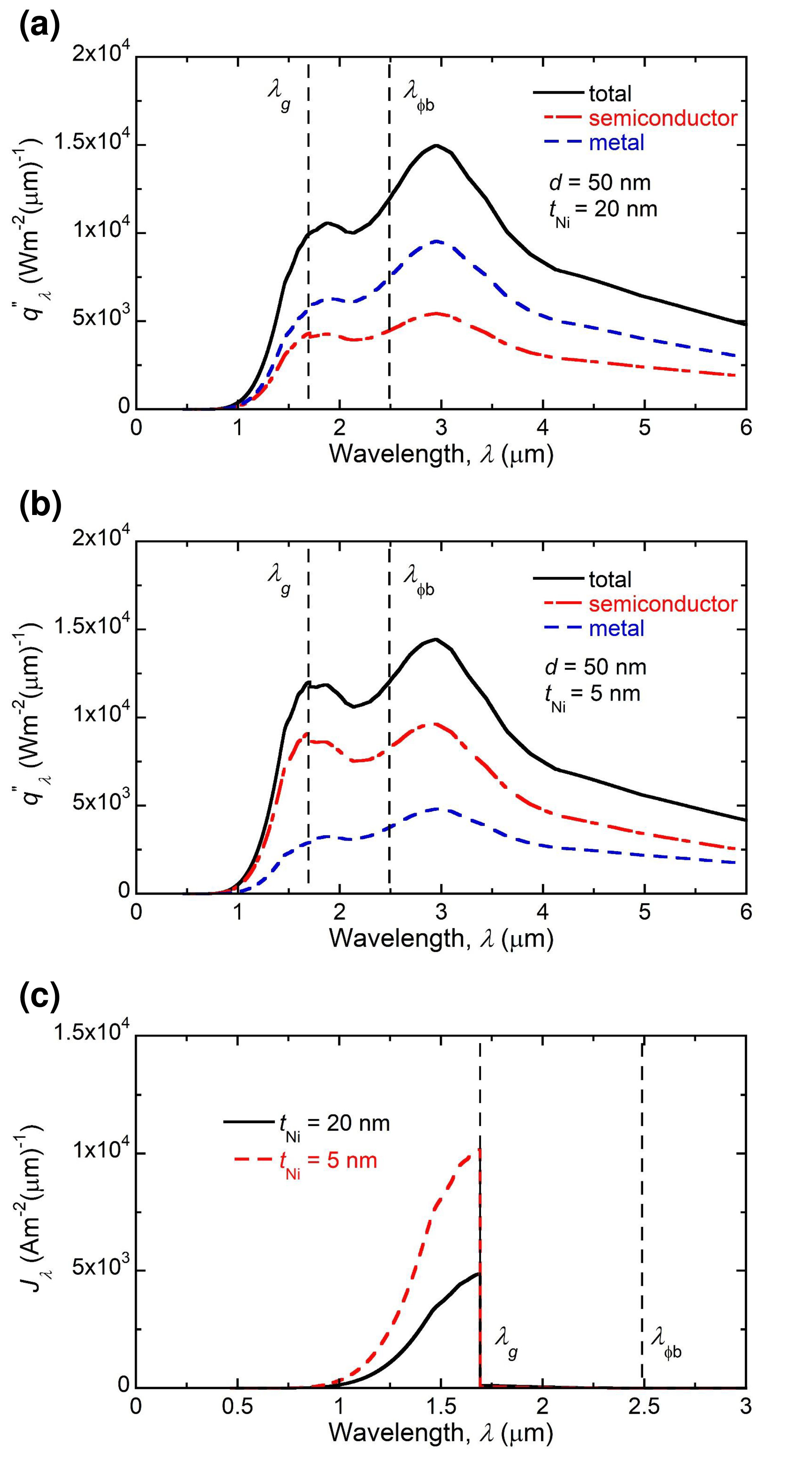}
\caption{(a-b) Spectral heat flux, $q^{\prime\prime}_{\lambda}$ when $t_{Ni}= 20$ nm and 5 nm, respectively. (c) Comparison of the net spectral photocurrent, $J_{\lambda,\text{net}}$ for both cases.}
\label{Fig:3}
\end{figure}
%%%%%%%

Figures \ref{Fig:3}a and \ref{Fig:3}b describe the spectral radiative heat flux absorbed in the metal and in the semiconductor when $t_{Ni}= 20$ nm and 5 nm, respectively at $d=50$ nm. When $t_{Ni}= 20$ nm in Fig.\ \ref{Fig:3}a, the absorbed radiative energy in the metal layer 1.6 times greater than that in the semiconductor. As $t_{Ni}$ decreases to 5 nm (see Fig.\ \ref{Fig:3}b), the absorption in the metal layer decreases, leading to only 35\% of the total radiative heat flux is absorbed by the metal layer.  Although the total radiative heat flux absorbed in both the metal film and the semiconductor decreases as $t_{Ni}$ decreases, the configuration with the 5-nm-thick nickel layer yields larger photocurrent (see Fig.\ \ref{Fig:3}c). This is because for the thinner Ni configuration, the photons whose energy is greater than the bandgap energy can be absorbed more in the semiconductor side (i.e., having high quantum efficiency) rather than in the metal side (i.e., having low quantum efficiency). By decreasing the thickness of Ni layer to 5 nm, the total photocurrent is increased by 2.2 times, and the maximum electrical power is enhanced by 2.7 times compared to the case with 20-nm-thick nickel layer. Furthermore, because the total radiative heat flux also decreases, the conversion efficiency (i.e., the maximum power output divided by the total radiative heat flux absorbed in the TPV cell) increases 2.8 times with the 5-nm-thick nickel layer.  

%%%%%%%%%%
\begin{figure}[!b]
\centering\includegraphics[width=0.95\textwidth]{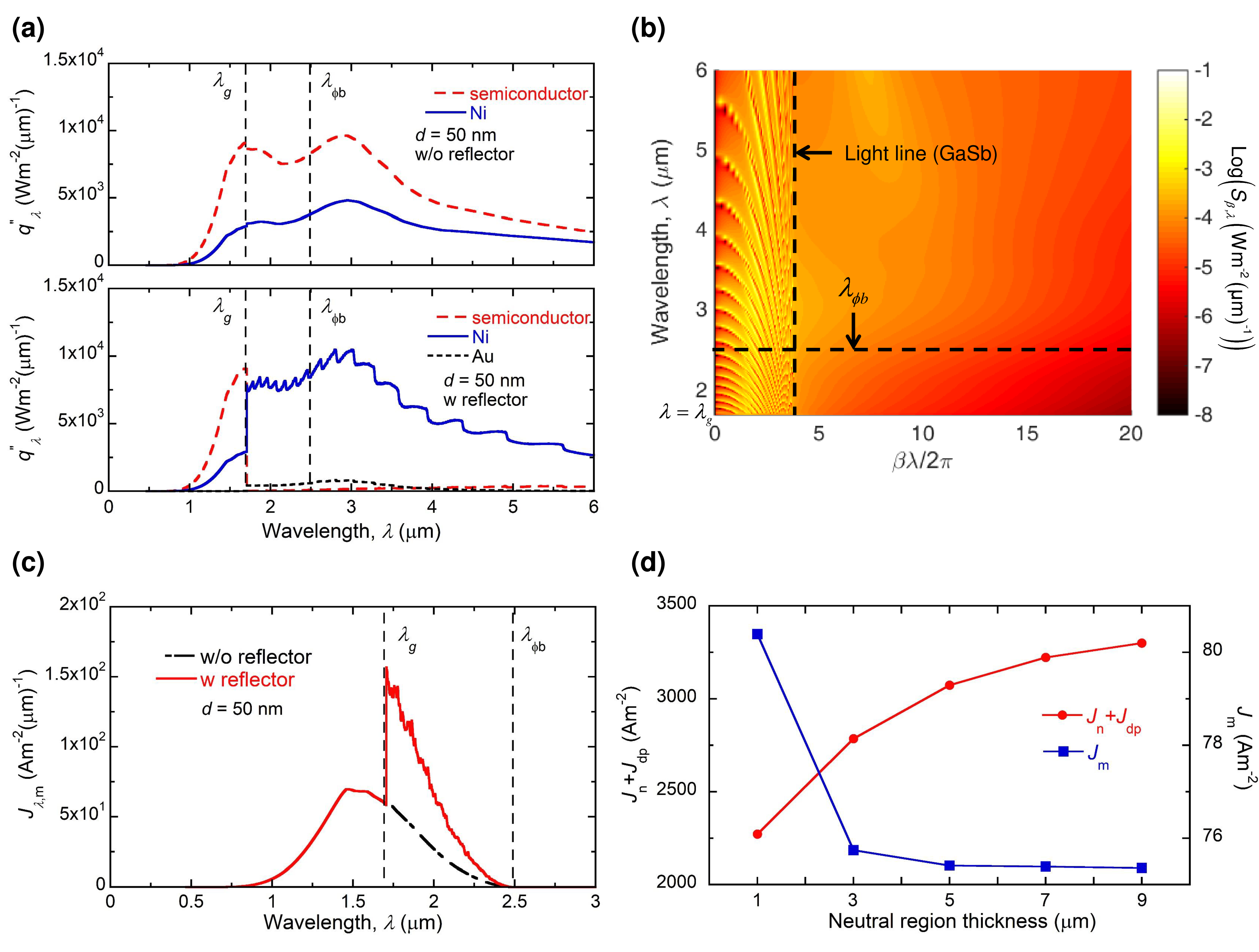}
\caption{(a) Spectral heat flux, $q^{\prime\prime}_{\lambda}$ without (upper panel) or with (lower panel) the backside reflector. (b) Contour plot of $S_{\beta,\lambda}(\beta,\lambda)$ absorbed in the nickel layer. The $x$-axis indicates the normalized parallel wavevector, and the vertically dashed line indicates the GaSb light line inside which wave propagates in the GaSb TPV cell at $0 < \beta\lambda/2\pi < n_{\text{GaSb}}$. (c) Spectral photocurrent, $J_{m,\lambda}$ with or without the backside reflector. (d) Total photocurrent density generated in the semiconductor ($J_n+J_{dp}$) and metal side ($J_m$) with respect to the thickness of neutral region.}
\label{Fig:4}
\end{figure}
%%%%%%%

It can be readily seen from from Figs. \ref{Fig:2} and \ref{Fig:3} that significant portion of the radiative heat flux absorbed in the semiconductor dose not produce the photocurrent. This can result in the rise of the cell temperature (i.e., thermal loss), which may increase the dark current and deteriorate the performance of near-field TPV system. In order to deal with the wasted heat absorbed in the $n$-doped GaSb, thin Au film that can form ohmic contact with GaSb can be introduced at the bottom of the TPV cell. To do so, the thickness of GaSb semiconductor is set to be 5 $\mu$m considering the penetration depth ($\sim1$ $\mu$m) of the spectral near-field radiative heat flux in $\lambda<\lambda_g$. Figure \ref{Fig:4}a describes the spectral heat flux absorbed in the semiconductor and in metal with or without the Au-backside reflector. Because Au has a high reflectivity in the infrared spectral region, photons that reach the Au-backside reflector (mostly $\lambda>\lambda_g$) can be effectively reflected back. Considering that GaSb rarely absorbs the radiative heat flux in the spectral region of $\lambda>\lambda_g$, the absorption in the finite-thickness GaSb layer in this wavelength interval is significantly reduced, as also noted in Refs.\ \cite{bright2014performance, lim2018optimization}. Interestingly, the absorbed energy for $\lambda>\lambda_g$ in the nickel metal layer significantly increases with the Au-backside reflector.

To elucidate the physical mechanism of the increased absorption in the Ni layer due to the Au-backside reflector, the contour of $S_{\beta,\lambda}$ absorbed in the nickel layer is plotted in Fig.\ \ref{Fig:4}b. Due to multiple reflections between Ni and Au, Fabry-Perot-like resonance modes appear inside the GaSb light line. The enhanced absorption due to the Fabry-Perot-like resonances will occur more densely in the wavelength-wavevector space for the thicker GaSb. Because the spectral radiative flux above the Schottky barrier height can be re-absorbed by the nickel layer, the resulting spectral internal emission photocurrent for $\lambda_g < \lambda <\lambda_{\phi_b}$ can increase significantly (refer to Fig.\ \ref{Fig:4}c), leading to the 1.5-times enhancement in $J_m$. However, the 1.5-times enhancement in $J_m$ by adding the backside reflector cannot enhance the power output significantly because the contribution of the internal emission photocurrent to the photocurrent generation is negligible. Moreover, both the spectral heat flux and the spectral photocurrent at $\lambda < \lambda_g$ will not change much because most of the radiative heat flux with the energy greater than the bandgap energy will be absorbed before reaching the Au reflector. Therefore, the Au-backside reflector results in the 1.1-times enhancement in the conversion efficiency mainly due to the reduction of the absorbed radiative heat flux in the semiconductor. 

To investigate effects of the thickness of the neutral region on the performance of the Schottky-junction-based near-field TPV system, the photocurrent density is calculated for different thicknesses of the neutral region in Fig.\ \ref{Fig:4}d. Here, the finite difference method is used to calculate the photocurrent generated in the finite-thickness semiconductor instead of the semi-analytic method. The boundary condition at the interface between the semiconductor and the Au reflector is $\left.D_h\frac{dn_h}{dz}\right|_{z=w_{dp}+w_n}=-u_n\left [ n_h(w_{dp}+w_n)-n^0_h\right ]$, where $u_n=10000$ m$\text{s}^{-1}$ is the surface recombination velocity \cite{vigil2005passivation} and $w_n$ is the thickness of the neutral region. It can be seen from Fig.\ \ref{Fig:4}d that the thinner neutral region results in the less generated photocurrent in the semiconductor side (i.e., $J_n+J_{dp}$). This is due to the increased minority carrier loss caused by the surface recombination at the GaSb-Au interface. If the thickness of the neutral region further decreases and becomes comparable to or smaller than the radiation penetration depth, the internal emission photocurrent can be increased because the radiative energy greater than the bandgap ($\lambda < \lambda_g$) can be reflected back and re-absorbed in the Ni layer. In fact, Fig.\ \ref{Fig:4}d clearly shows that $J_m$, which was nearly constant in $3\,\mu\text{m}< w_n < 9\,\mu\text{m}$, is significantly improved as $w_n$ decreases to 1 $\mu$m. However, because the increase of $J_m$ is much smaller than the reduction of $J_n+J_{dp}$, reducing $w_n$ is not advantageous if the quantum efficiency for the internal emission photocurrent is extremely low.

According to Ref.\ \cite{scales2010thin}, if one employ an extremely thin metal layer (i.e., thinner than 5 nm considered here) using materials whose electron mean free path is longer than nickel, the quantum efficiency of the internal emission photocurrent can be improved. As a limiting case, when the ratio of the electron mean free path to the thickness of metal layer becomes infinite, all electrons having energy greater than $\phi_b$ can surmount the Schottky barrier height, resulting in the quantum efficiency $\eta_{m,\lambda}=1-\frac{\lambda\phi_b}{hc_0}$. If we assume such a limiting case to the system with 1-$\mu$m-thick neutral region, the internal emission photocurrent (2136 Am$^{-2}$) can be the same order of magnitude to the diffusion photocurrent (1852 Am$^{-2}$) by introducing the Au reflector, leading to 1.37 time enhancement in the power output and 1.81 time enhancement in the conversion efficiency. In addition, previous studies suggested that the quantum efficiency for the internal emission photocurrent can overcome the limit of $\eta_{\lambda,m}=1-\frac{\lambda\phi_b}{hc_0}$ if hot carrier is generated by surface plasmon polariton (SPP) \cite{bernardi2015theory, sundararaman2014theoretical}. The SPP supported in the thin metal can be coupled to the SPP supported in the emitter, which may lead to significant enhancement of the absorbed radiation in both the metal side and the semiconductor side of the TPV cell. Therefore, considering all types of photocurrent generation in the Schottky-junction near-field TPV system is crucial in designing the system with high performance.

%%==========================================
\section{Conclusions}
The performance of Schottky-junction-based near-field TPV system was investigated with the detailed model considering photocurrent generation in the semiconductor side of TPV cell. When the vacuum gap distance between the tungsten emitter and the nickel-$n$-doped GaSb Schottky-junction-based TPV cell is 50 nm, the ratio of generated diffusion, drift, and internal emission photocurrents was found to be 1 : 0.12 : 0.016, meaning that the photocurrent generation in the semiconductor side takes the most of the photocurrent generation in the Schottky-junction-based TPV cell. Resultant power output is 253 Wm$^{-2}$, which is 24 times greater than that of the far-field TPV system. It was found that depositing the thinner nickel layer is advantageous in enhancing the performance of near-field TPV system. Further, it was shown that by introducing the backside reflector, the radiative heat flux with energy greater than the Schottky barrier height can be re-absorbed in the Ni layer, and the corresponding internal emission photocurrent increases 1.5 times. Although the Au-backside reflector cannot enhance the power output largely, it can enhance the conversion efficiency by reducing the waste heat absorption in the semiconductor side. In order to obtain the remarkable enhancement in the power output with the backside reflector, the quantum efficiency of the internal emission photocurrent should be improved first by modifying the metal geometry or employing the hot-carrier generation accompanied with the SPP. The quantitative study presented here shows that the role of the semiconductor is crucial in analyzing the performance of the Schottky-junction-based near-field TPV system and will promote future study and implementation of near-field TPV system.\\

\begin{acknowledgments}
This research was supported by the Creative Materials Discovery Program (NRF-2018M3D1A1058972) as well as the Basic Science Research Program (NRF-2017R1A2B2011192) through the National Research Foundation of Korea (NRF) funded by Ministry of Science and ICT. 
\end{acknowledgments}

\bibliography{Song_Bib}

\end{document}